\documentclass[aps,prb,showpacs,twocolumn]{revtex4}
\usepackage{amsfonts}
\usepackage{amssymb}
\usepackage{amsmath}
\usepackage{graphicx}
\usepackage{verbatim}

\begin{document}

\title{Hubbard model description of silicon spin qubits: charge stability diagram and tunnel coupling in Si double quantum dots}
\author{S. Das Sarma, Xin Wang, and Shuo Yang}
\affiliation{Condensed Matter Theory Center, Department of Physics,
University of Maryland, College Park, MD 20742}

\begin{abstract}
We apply the recently introduced Hubbard model approach to quantitatively describe the experimental charge stability diagram and tunnel coupling of silicon double quantum dot systems. The results calculated from both the generalized Hubbard model and the microscopic theory are compared with existing experimental data, and excellent agreement between theory and experiment is found. The central approximation of our theory is a reduction of the full multi-electron multi-band system to an effective two-electron model, which is numerically tractable. In the microscopic theory we utilize the Hund-Mulliken approximation to the electron wave functions and compare the results calculated with two different forms of confinement potentials (biquadratic and Gaussian).
We discuss the implications of our work for future studies.
\end{abstract}


\maketitle

\section{Introduction}

Semiconductor solid state structures are believed to be the most promising prospective hardware for building a quantum computer.\cite{
Loss.98,DiVincenzo.00,Levy.02,Koppens.06} Their decisive advantage, based on the existence of the huge semiconductor electronics technology, 
is their scalability\cite{Kane.98,Vrijen.00,Friesen.03,Taylor.05}: once few-qubit operations are successfully accomplished, the well-established
 modern semiconductor microelectronics industry greatly facilitates scaling up to many qubits. Although long coherence times have been demonstrated in both GaAs\cite{Kikkawa.98,Amasha.08,Koppens.08,Barthel.10,Bluhm.11} and Si\cite{Tyryshkin.03,Morello.10,Simmons.10,Xiao.10a} semiconductor devices, the two architectures are at different stages of development en route to an eventually viable quantum computer. In the GaAs quantum dot system, one- and two-qubit manipulations have been experimentally demonstrated.\cite{Petta.05,Koppens.06,Foletti.09,Laird.10,vanWeperen.11} However, in its counterpart silicon-based system, these manipulations are yet to be satisfactorily achieved, in spite of much efforts in several laboratories across the world.

Several features distinguish the silicon-based devices from the GaAs systems. First, the silicon quantum dot systems generally have longer spin coherence time than the GaAs system, due to some novel properties such as weak spin-orbit coupling and low nuclear spin density.\cite{Tyryshkin.03, Tahan.02, deSousa.03, Fedichkin.04, Appelbaum.07} In this regard, silicon-based devices are often considered to be ideal for quantum computation, because quantum error corrections protocols are easier to implement for qubits with longer coherence times. However, various traps and impurities introduced during the fabrication lead to unpredictable local fluctuations of the confinement potential, which has so far prevented a coherent manipulation in these quantum dot systems.\cite{Nordberg.09}
In this regard, the Si quantum dot qubit is inferior to the GaAs quantum dot structure since the interface quality in the Si system, particularly for the Si/SiO$_2$ system (but also to a lesser degree for the Si/SiGe system), is substantially worse than the corresponding epitaxially perfect GaAs/AlGaAs system.
 Second, the valley degree of freedom resulting from the special band structure of silicon, with bulk Si having six equivalent conduction band minima (i.e. a valley degeneracy of six), further complicates the problem.\cite{dassarma.05,LiQZ.10,Culcer.10, Goswami.07,Borselli.11}  We also note that in silicon metal-oxide-semiconductor  
(Si-MOS) system as well as the SiGe system the Coulomb interactions are stronger than that of the GaAs devices, due to lower dielectric constant and larger effective mass, which consequently affects the exchange interaction.\cite{LiQZ.10} The larger electron effective mass of Si compared with GaAs makes the quantum tunneling weaker and the effective Bohr radius smaller, leading to tighter lithographic constraints in Si over GaAs for achieving equivalent qubit properties, thus rendering the fabrication of Si qubits much more difficult experimentally.

In spite of these experimental difficulties, considerable progress has recently been made experimentally in silicon based systems, including both Si-MOS devices\cite{HuB.09,Nordberg.09,Lim.09b,Xiao.10b}
and Si/SiGe structures.\cite{Simmons.10,Goswami.07,Klein.07,Simmons.07,Shaji.08,
Simmons.10prb,Borselli.11} Particularly important in the few-electron regime of a double quantum dot system is the charge stability diagram\cite{Liu.08a,Pierre.09} (explained in more detail below), which is an intuitive and physical manifestation of the Coulomb blockade\cite{Livermore.96} physics as well as the controllability of the qubit via, for example, the tunnel coupling between the two dots. The variations of the tunnel coupling not only affect the exchange interaction in the actual qubit manipulation in a chosen subspace of the charge stability diagram, but also lead to a change in the geometry of the overall charge stability diagram. Lai \emph{et al.}\cite{Lai.10} and Simmons \emph{et al.}\cite{Simmons.09} have demonstrated highly tunable double quantum dot systems on Si-MOS and Si/SiGe systems, respectively. The charge stability diagrams over a wide range of electron occupancy were obtained in these experiments, and most interestingly they have shown how the patterns of the stability diagrams change as the tunnel coupling between the two dots is varied. In the experiment on Si/SiGe\cite{Simmons.09} the numerical value of the tunnel coupling is extracted from experiments using a two-level model.\cite{DiCarlo.04}
These experiments have paved the way for the realization of singlet/triplet qubits in silicon, and a quantitative understanding of quantum effects in the charge stability diagrams in Si coupled quantum dot experiments is the central goal of the present paper. 

In a previous publication\cite{Yang.11} we have advocated a quantum theory using the generalized Hubbard model to describe the double quantum dot system.
The advantage of our theory over the traditionally used capacitance circuit model\cite{Wiel.03,Hanson.07} is that our model naturally includes quantum fluctuations, which is particularly suitable for the discussion of the physics of the tunnel coupling in the context of charge stability diagrams.
Although our approach is completely general since the only restrictions are symmetry considerations, we focused on the GaAs system in our original\cite{Yang.11} work. In this paper we apply our method to the silicon quantum dot system. 
We start with a qualitative survey of the experimental results, followed by a detailed explanation of the assumptions and simplifications that we have made in applying our general theory to the Si double quantum dot system. Then we demonstrate that upon appropriately choosing the parameters, our theory is capable of giving quantitative description of the experimental charge stability diagram and tunnel coupling in the Si system. We also outline several future research directions implied by our work.

The remainder of the paper is organized as follows. In Sec.~\ref{sec:qualitative} we briefly discuss experimental results and assumptions that have been employed. In Sec.~\ref{sec:methods} we explain the generalized Hubbard model and the microscopic theory used in this work. Sec.~\ref{sec:results} presents calculated results on the charge stability diagram for both Si-MOS and Si/SiGe system, and the tunnel coupling in the Si/SiGe system. The results are compared to experiments side-by-side. 
Sec.~\ref{sec:conculsion} includes our discussion and conclusions.

\section{Qualitative considerations}\label{sec:qualitative}

In the typical experimental setup of Ref.~\onlinecite{Lai.10}, the double quantum dots are confined by three barrier gates (left, central and right) and the depth of the potential wells in the dots is controlled by two plunger gates with voltages $V_{P1}$ and $V_{P2}$ respectively, in analogy to the left ($V_L$) and right ($V_R$) gate voltages of GaAs devices.\cite{Petta.05} To shrink the dot size, the plunger gates are eliminated in Ref.~\onlinecite{Simmons.09}, and the gate voltages $V_L$ and $V_R$ are used instead. The Coulomb blockade\cite{Livermore.96} means that the electron occupancy in each dot (denoted by $n_i$ for dot $i$, $i=1,2$) is typically changed by an integer as the gate voltages are swept, which can be directly detected by measuring the tunneling current, using a quantum point contact.\cite{Elzerman.03}  As a consequence, a plot of the differential conductance on the plane with $V_{P1}$ and $V_{P2}$ (or $V_L$ and $V_R$) as axes is a visualization of electron configuration in the quantum dot system and is therefore called the charge stability diagram.
Fig.~\ref{fitLai}(a)-(c) show the charge stability diagrams measured by Lai \emph{et al.} on Si-MOS devices, reproduced from Fig.~2(a)-(c) of Ref.~\onlinecite{Lai.10}. The white regimes indicate the Coulomb blockade plateaus in which some particular set of electron occupancy $(n_1,n_2)$ is maintained; the colored lines separating the white regimes represent the electron tunneling changing the occupancy by integer numbers. Fig.~\ref{fitSimmons}(a)-(c) show similar results in Si/SiGe structures, reproduced from Fig.~2(c)-(a) of Ref.~\onlinecite{Simmons.09}. [Note that the order in the two sets of figures is reversed here to comply with the convention that inter-dot coupling increases from panel (a) to panel (c).]

The tunnel coupling between the two quantum dots can be adjusted by changing the electrostatic potential of the central barrier gate, which consequently affects the geometry of the charge stability diagram. The variation from panels (a) to (c) of Figs.~\ref{fitLai} and \ref{fitSimmons} illustrates this process, whose qualitative feature is well understood.\cite{Wiel.03,Hanson.07} If the central barrier is high enough to well separate the two dots (so that the coupling between them is weak), the two dots behave basically independently and the charge stability diagram has checker board pattern [Fig.~\ref{fitLai}(a)]. As the central barrier is lowered the inter-dot coupling becomes intermediate, the charge stability diagram becomes honeycomb-shaped [Fig.~\ref{fitLai}(b), Fig.~\ref{fitSimmons}(a) and (b)]. In Fig.~\ref{fitLai}(c) and Fig.~\ref{fitSimmons}(c) the inter-dot coupling is strong due to further lowering of the central barrier such that the two dots effectively merge into one, the stability diagram becomes like a series of diagonal parallel lines separating regimes with different total occupancy $N=n_1+n_2$, while the boundaries between different $(n_1,n_2)$ states within the same $N$ block [e.g. $(n_1+1,n_2)\rightarrow(n_1,n_2+1)$] are hardly visible due to the hybridization from the quantum fluctuations. In this situation with inter-dot tunneling induced strong quantum fluctuations, the classical assumption of integer electron filling per dot becomes invalid, thus making the classical circuit model inapplicable.

The prevailing quantitative approach of understanding the charge stability diagram is the capacitance circuit model, which symbolizes the Coulomb interaction by capacitors.\cite{Wiel.03} In our previous work \cite{Yang.11} we have shown a mapping of the capacitance circuit model to the specific limit of the generalized Hubbard model with quantum fluctuations completely suppressed.
In Fig.~\ref{fitLai}(a), the inter-dot coupling including both Coulomb interaction and the quantum fluctuations is small, and the capacitance model is, to a good approximation, valid. However,
in Fig.~\ref{fitLai}(b) the boundaries separating different states are rounded near the triple points due to quantum effects, while in Fig.~\ref{fitLai}(c) the quantum effects (rounding) become even more evident. The rounding feature of boundary lines cannot even be qualitatively explained in the capacitance model in any way and the basic purpose of the present paper is to give a quantitative description of the experiments using our proposed theoretical approach, combining the generalized Hubbard model and the microscopic theory.

In principle, a detailed quantitative theoretical description of a specific experiment needs to incorporate as many experimental parameters as possible extracted directly from the data. However in our situation several complications arise. These Si quantum dot experiments are done in the multi-electron regime, and the exact electron occupancy $(n_1,n_2)$ is unknown. While the capacitance model predicts a periodic form of the charge stability diagram, the actual situation is far from clear (and does not seem periodic). A realistic multi-electron and multi-band calculation would be extremely time-consuming and completely impractical.
Therefore we shall focus on the ``effective'' two-electron regime, meaning that we select one diamond in each stability diagram [indicated as blue square frames in Fig.~\ref{fitLai}(a)-(c)], and argue that the electron occupancies in the regime enclosed by the frame ranging from $(n_1,n_2)$ to $(n_1+2,n_2+2)$ are approximated by effective electron occupancies ranging from $(0,0)$ to $(2,2)$, although in reality the actual configuration is changing from $(n_1,n_2)$ to $(n_1+2,n_2+2)$ with both $n_1$ and $n_2$ being unknown (and possibly of order 10-20). We shall further assume that the multi-electron multi-band effect can be safely regarded as a renormalization of the effective two-electron and one-band Hubbard model parameters such as the Coulomb interaction $U$ but otherwise the same physics remains.\cite{long} Indeed a close examination of the experimental plots Fig.~\ref{fitLai}(a)-(c) shows that the stability diagrams are, to a good approximation, periodic over a wide range, with details varying slightly as the occupancies are changed. This indicates that our assumptions are reasonable.
A complete clarification of this multi-electron issue requires further experimental investigation on a much wider range of the stability diagram, new experimental techniques that enable precise measurement of electron occupancies, and a substantially more complicated theoretical and computational study on a multi-band system. This is far beyond the scope of the present paper and probably the current experiments.

Another consequence of the aforementioned approximation is that some of the gate voltages  (for example the plunger gate voltages $V_{P1}$ and $V_{P2}$) are no longer basic variables in our theory since they control all electrons while electrons in the effective two-electron regime simply feel a different potential which depends on the overall electrostatic potential, the separation between the fabricated gate and the actual localized electron gas, and the underlying Fermi sea of the background two-dimensional electron gas in the semiconductor heterostructure. To facilitate the discussion we adopt the language that is used in the study of GaAs devices, i.e. the left and right gate voltages $V_L$ and $V_R$ as the effective gate voltages for Si-MOS devices.
We will not try to establish a precise connection between $(V_{P2},V_{P1})$ and $(V_R,V_L)$ partly because one has no reliable information of the electrons in the underlying Fermi sea and partly because the fabrication-induced disorder may come into play. We shall, however, assume that $(V_R,V_L)$ are related to $(V_{P2},V_{P1})$ linearly, i.e. the calculated stability diagram should fit the experimental plot after appropriate rescaling and shifting. In fact, we have slightly rescaled the $y$-direction of the experimental plot of Ref.~\onlinecite{Lai.10} in Fig.~\ref{fitLai}(a)-(c) since according to our previous work\cite{Yang.11} the boundaries separating different states within the same $N$ block [for example the one separating $(1,1)$ and $(0,2)$ regime] should be at $45^\circ$ with respect to the $x$-axis (i.e. parallel to the line $V_L=V_R$) if the mapping to the capacitance model is valid. The experimental plots shown in Ref.~\onlinecite{Lai.10} need be compressed along the $y$-direction in order to exhibit this property. This rescaling is simply a matter of convenience and convention, and does not affect any of our results or conclusions.

In addition to the assumptions above, we further assume that the Si conduction band valley splitting is sufficiently large at the interface so that the valley degree of freedom can be disregarded. The neglecting of the valley splitting because of interface effects is the standard approximation in essentially all of the Si spin qubit literature with only a few exceptions.\cite{Culcer.10,Culcer.10b} We focus our study on the case without any external magnetic field, but the generalization to the finite magnetic field situation\cite{long} is straightforward.

\section{Methods}\label{sec:methods}

In this section we discuss the methods used to model and quantify the experiments. We first describe the generalized Hubbard model, and then the microscopic theory where we use two different models of  confinement potentials since the actual quantum dot confinement potential is unknown. We also briefly outline how we fix the parameters to best describe the experiments.

In our previous publication,\cite{Yang.11} we have proposed a generalized form of Hubbard model, retaining all terms allowed by symmetry (total particle number $N$ and total spin $S_z$ conserved). The Hamiltonian can be expressed as
\begin{equation}
H=H_\mu+H_t+H_U+H_J.
\end{equation}
The one-body part includes the chemical potential/level energy terms
\begin{equation}
H_\mu=-\sum_{i\sigma}\mu_in_{i\sigma},\end{equation}
and the tunnel coupling (hopping) terms
\begin{equation}
H_t=-\sum_{\sigma}\left(tc_{1\sigma}^\dagger c_{2\sigma}+h.c.\right),
\end{equation}
where $n_{i\sigma}=c_{i\sigma}^\dagger c_{i\sigma}$, $\mu_i$  is the chemical potential of electrons on site $i$ ($i=1,2$), and $t$ is the tunnel coupling or inter-site hopping.

The two-body part consists of a direct Coulomb repulsion term $H_U$
\begin{equation}
\begin{split}
H_U&=U_1n_{1\uparrow}n_{1\downarrow}+U_2n_{2\uparrow}n_{2\downarrow}+U_{12}(n_{1\uparrow}n_{2\downarrow}+n_{1\downarrow}n_{2\uparrow})\\
&+(U_{12}-J_e)(n_{1\uparrow}n_{2\uparrow}+n_{1\downarrow}n_{2\downarrow}),\end{split}
\end{equation} 
and a term $H_J$ including spin-exchange ($J_e$), pair-hopping ($J_p$), and occupation-modulated hopping terms ($J_t$)
\begin{equation}
\begin{split}
H_J&=-J_ec_{1\downarrow}^\dagger c_{2\uparrow}^\dagger c_{2\downarrow}c_{1\uparrow}-J_pc_{2\uparrow}^\dagger c_{2\downarrow}^\dagger c_{1\uparrow}c_{1\downarrow}\\
&-\sum_{i\sigma}J_{ti}n_{i\sigma}c_{1\overline{\sigma}}^\dagger c_{2\overline{\sigma}}+h.c..\end{split}
\end{equation} 

The gate voltages $V_L$ and $V_R$ appear in the generalized Hubbard model in terms of the chemical potentials $\mu_{1,2}$ with the relation\cite{Yang.11,Gaudreau.06,Korkusinski.07}
\begin{align}
\mu_{1}&=\left| e \right| \left[ \alpha_{1} V_{L} + \left( 1- \alpha_{1} \right) V_{R} \right] + \gamma_{1}; \notag \\
\mu_{2}&=\left| e \right| \left[ \left( 1- \alpha_{2} \right) V_{L} + \alpha_{2} V_{R} \right] + \gamma_{2}.
\end{align}
If the mapping to the capacitance model\cite{Yang.11} is valid, we have
\begin{align}
\alpha_{1} & = \frac{\left(U_{2}-U_{12}\right)U_{1}}{U_{1}U_{2}-U_{12}^{2}},\quad\alpha_{2}=\frac{\left(U_{1}-U_{12}\right)U_{2}}{U_{1}U_{2}-U_{12}^2}.
\label{alphabeta}
\end{align}
We note that when the quantum fluctuations are noticeable such that the capacitance model does not apply, one is not necessarily restricted to Eq.~\eqref{alphabeta} to convert the result from $(\mu_2,\mu_1)$ to $(V_R,V_L)$ plane. For simplicity, in this work we shall only use the Coulomb parameters $U_1$, $U_2$, $U_{12}$ and the tunnel coupling  $t$ to fit the experiments, assuming other exchange parameters to be small, although they are invariably present in both microscopic calculations and actual experiments. Our motivation at this initial stage of the development of the subject is to keep as few free parameters as possible in the theory and still incorporate quantum fluctuation effects not accessible in the widely used capacitance circuit models.

As discussed in Ref.~\onlinecite{Yang.11}, the generalized Hubbard model must be backed up by a microscopic theory.\cite{Burkard.99,Hu.00,Hu.01,Sousa.01,LiQZ.10,Gimenez.07,Nielsen.10} The many-electron Schr\"odinger equation is solved in the presence of a double-well confinement potential. In each well the potential is approximated as a harmonic oscillator whose low-lying eigenstates are Fock-Darwin states.
According to the Hund-Mulliken approach, the two-electron wave functions are constructed by hybridizing and orthogonalizing the Fock-Darwin states. These two-electron wave functions span a Hilbert space from which all the parameters of the generalized Hubbard model are derived. The details are described in Ref.~\onlinecite{Yang.11}. Here, we use two different kinds of potentials: a biquadratic form [Eq.~\eqref{biquadratic}] and a Gaussian form [Eq.~\eqref{gaussian}], and make a side-by-side comparison of the results. We start with the biquadratic potential,\cite{LiQZ.10, Helle.05, Pedersen.07}
\begin{equation}
\begin{split}
V_Q(x,y)={\rm Min}\Big\{& \frac{m \omega_1^{2}}{2} [(x+a)^{2}+y^{2}]-\mu_{1},\\
\frac{m \omega_2^{2}}{2} & [(x-a)^{2}+y^{2}]-\mu_{2}, 0 \Big\}.
\end{split}\label{biquadratic}
\end{equation}

The biquadratic potential has a simple form which only involves three parameters: $\omega_1$, $\omega_2$ and the inter-dot distance $2a$. We expect that the on-site Coulomb interactions $U_1$ and $U_2$ are directly related to $\omega_1$ and $\omega_2$, while the inter-dot Coulomb interaction should vary primarily with the inter-dot distance $2a$ with possible minor dependence on $\omega_1$ and $\omega_2$. This follows from the ``local'' assumption of the electron wave function that the electrons are well localized in their own respective potential wells.
This assumption is built into the Hund-Mulliken approximation, where the many-body wave functions are constructed using local wave functions. 
 (The constructed many-body wave functions are the true ones when the two potential wells are far from each other, or the central potential barrier is high, such that the two dots behave independently.) 
To understand this better we consider an isolated single quantum dot with harmonic confinement potential
\begin{align}
V_i\left(\boldsymbol{r}\right)=\frac{1}{2}m\omega_i^{2}\left[\left(x-x_{i}\right)^{2}+\left(y-y_{i}\right)^{2}\right]. 
\end{align}
The subscript $i=1,2$ labels the dots.
If the lowest orbital is occupied by two electrons with opposite spins,
the Coulomb interaction between them (i.e. the two electrons in the same dot labeled by ``$i$'') is
\begin{align}
U_i&=\int d\boldsymbol{r}_{1}\int d\boldsymbol{r}_{2}\left|\varphi_i\left(\boldsymbol{r}_{1}\right)\right|^{2}\left|\varphi_i\left(\boldsymbol{r}_{2}\right)\right|^{2}\frac{ke^{2}}{\left|\boldsymbol{r}_{1}-\boldsymbol{r}_{2}\right|}\notag\\
&=ke^2\sqrt{\frac{\pi}{2}}\frac{1}{l_{i}}=ke^2\sqrt{\frac{\pi m\omega_i}{2\hbar}}
\label{U1U2}
\end{align}
where $k=1/(4\pi\varepsilon_0\varepsilon)$,
the $S$-orbital wave function is
\begin{align}
\varphi_i\left(\boldsymbol{r}\right)=\frac{1}{\sqrt{\pi} l_{i}}\exp\left[-\frac{\left(x-x_{i}\right)^{2}+\left(y-y_{i}\right)^{2}}{2l_{i}^{2}}\right], 
\end{align}
and the effective length $l_{i}=\sqrt{\hbar/\left(m\omega_i\right)}$.
Therefore, the on-site Coulomb interactions only depend on $\omega_i$ in the isolated dot limit.
If we ignore the overlap between Fock-Darwin states of different dots,
the inter-site Coulomb interaction is
\begin{align}
U_{12} & = \int d\boldsymbol{r}_{1}\int d\boldsymbol{r}_{2}\left|\varphi_{1}\left(\boldsymbol{r}_{1}\right)\right|^{2}\left|\varphi_{2}\left(\boldsymbol{r}_{2}\right)\right|^{2}\frac{ke^{2}}{\left|\boldsymbol{r}_{1}-\boldsymbol{r}_{2}\right|} \notag \\
 & = ke^{2}\sqrt{\frac{\pi}{l_1^2+l_2^2}}\exp\left[-\frac{2a^{2}}{l_{1}^{2}+l_{2}^{2}}\right]{\rm I}_{0}\left(\frac{2a^{2}}{l_{1}^{2}+l_{2}^{2}}\right).
\label{U12}
\end{align}
Here $2a$ is the distance between the two dots, and ${\rm I}_0$ is the zeroth-order modified Bessel function. We see that $U_{12}$ is primarily determined by $a$.

These considerations make the fit using the biquadratic potential straightforward, which we outline taking Si-MOS as an example. For the case of Fig.~\ref{fitLai}(a) the two dots can be regarded as basically independent so $\omega_{1}$ and $\omega_{2}$ are determined from the Hubbard parameters $U_1$ and $U_2$ using Eq.~\eqref{U1U2}. From an appropriate $U_{12}$, the inter-dot distance $2a$ is determined by Eq.~\eqref{U12}. For the cases of Fig.~\ref{fitLai}(b) and (c), Eqs.~\eqref{U1U2} and \eqref{U12} provide a initial guess from which fine tuning takes place. One key feature of the biquadratic confinement model, making it preferable for our theory, is that the minimal set of three parameters entering our generalized Hubbard model (i.e. $U_1$, $U_2$, $U_{12}$) are determined by exactly three parameters ($\omega_1$, $\omega_2$, $a$) of the underlying microscopic confinement model. Attention must be paid when transforming from the $(\mu_2,\mu_1)$ to $(V_R,V_L)$ planes: the capacitance model is not valid in the cases of Fig.~\ref{fitLai}(b) and (c), therefore we are not restricted to the restriction Eq.~\eqref{alphabeta}, although we try to choose $(\alpha_1,\alpha_2)$ close to that predicted by the capacitance model.

We also do the microscopic calculation with the Gaussian potential,\cite{Hu.00}
\begin{equation}
\begin{split}
V_{G}(x,y)&=-\mu_{1} \exp \left[ -\frac{\left( x+a_1 \right)^{2} +y^{2}}{l_{d1}^{2}} \right]\\
-\mu_{2} \exp & \left[  -\frac{\left( x-a_2 \right)^{2} +y^{2}}{l_{d2}^{2}} \right] + V_{b} \exp \left[ -\frac{x^{2}}{l_{bx}^2} -\frac{y^{2}}{l_{by}^2} \right].
\end{split}\label{gaussian}
\end{equation}
The Gaussian potential involves seven independent parameters. $a_1$ and $a_2$ denote the distance from the center of the quantum dot to the middle potential barrier. $l_{d1}$ and $l_{d2}$ represent the size of the individual potential well. $V_b$ is the height of the middle potential barrier whose size is determined by $l_{bx}$ and $l_{by}$ in the $x$- and $y$-direction, respectively. Although simple analytical formulas connecting these microscopic parameters and the Hubbard model do not exist (particularly since the Hubbard model is characterized by three parameters and the Gaussian confinement model by seven parameters), the similar roles played by some parameters in the Gaussian model and the biquadratic model simplify the search. The parameters $a_1$ and $a_2$ in Eq.~\eqref{gaussian} should have similar order of magnitude as $a$ in Eq.~\eqref{biquadratic}, while $l_{d1}$ and $l_{d2}$ for the Gaussian potential are closely related to $\omega_1$ and $\omega_2$ in the biquadratic potential. After these parameters are appropriately chosen, both the height and the size of the central potential barrier are fine tuned to fit the experimental plots. 

The appropriate dielectric constant $\varepsilon$ (7.8 for Si-MOS and 12.375 for Si/SiGe) and the effective mass ($m^*=0.191m_e$) are used throughout the microscopic calculations.

\section{Results}\label{sec:results}

\subsection{Si-MOS}

\begin{figure*}
    \centering
    \includegraphics[width=13.5cm]{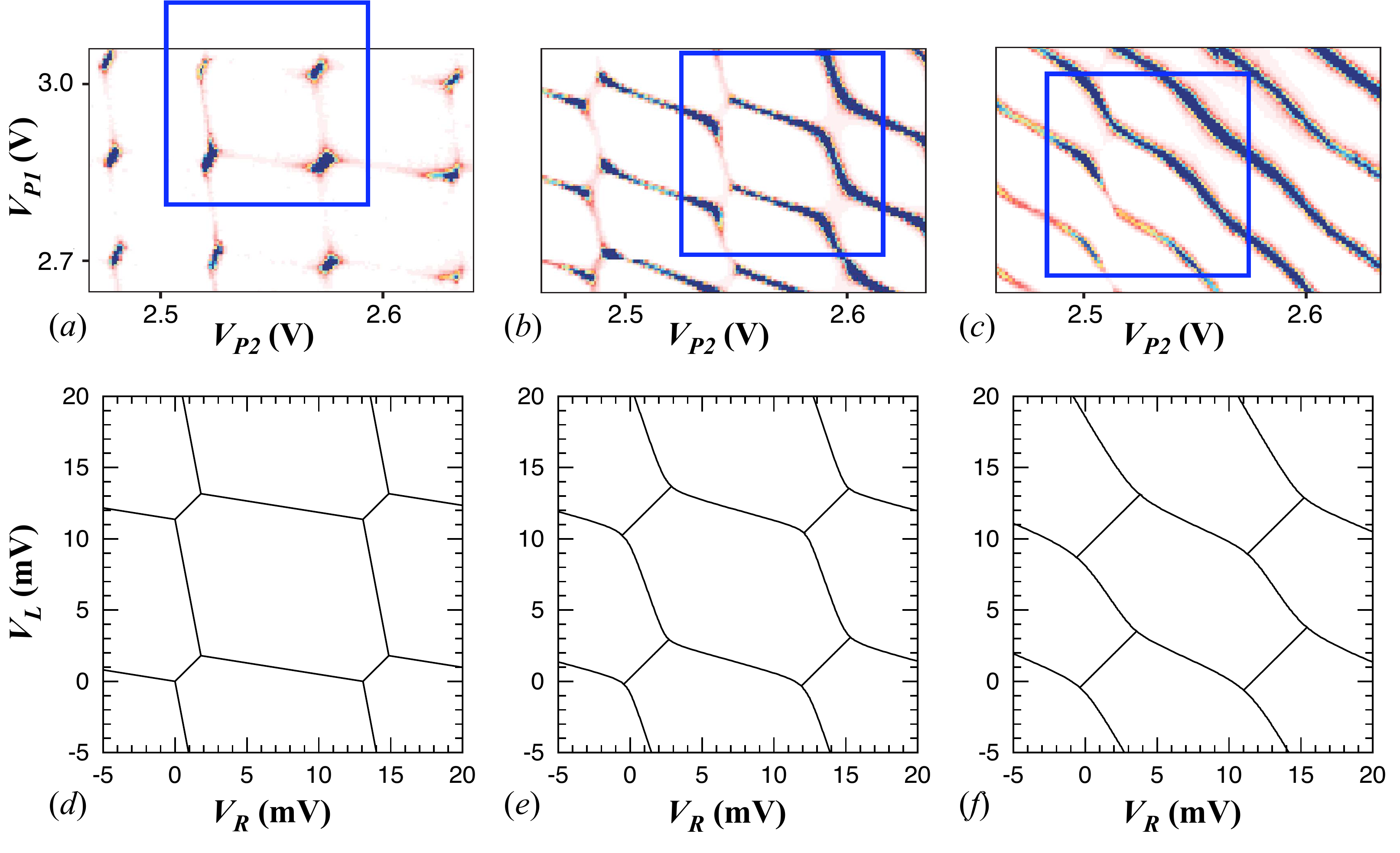}
\includegraphics[width=13.56cm]{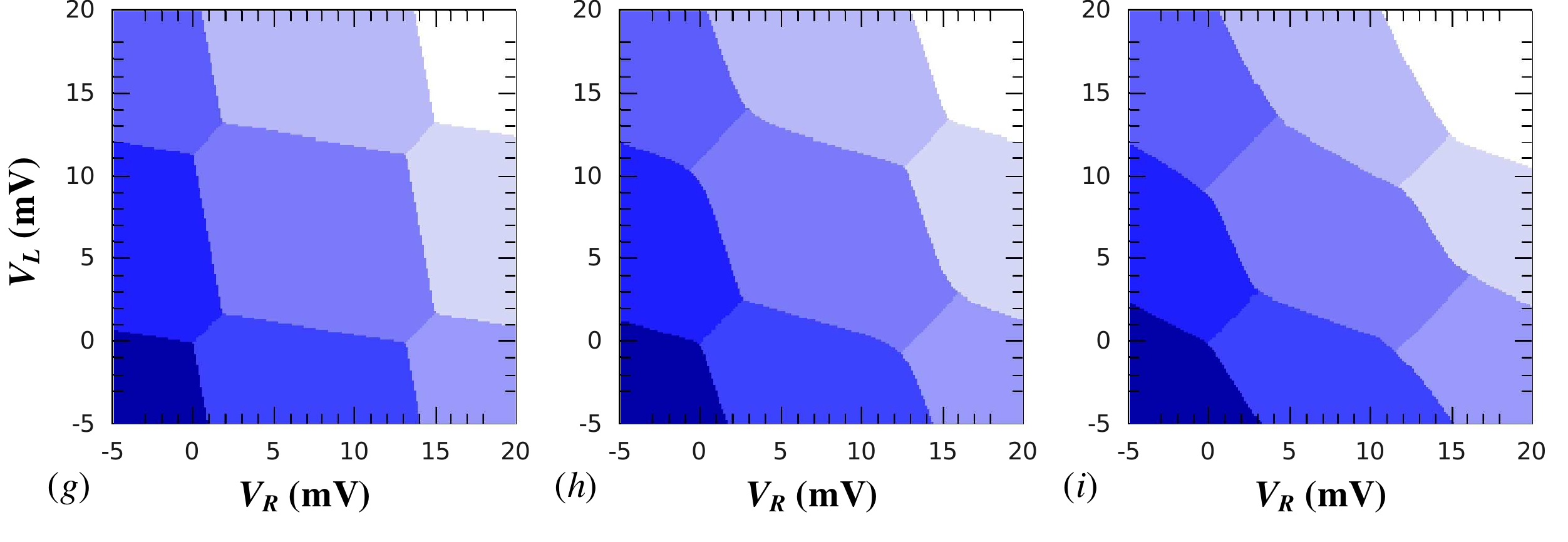}
\includegraphics[width=13.56cm]{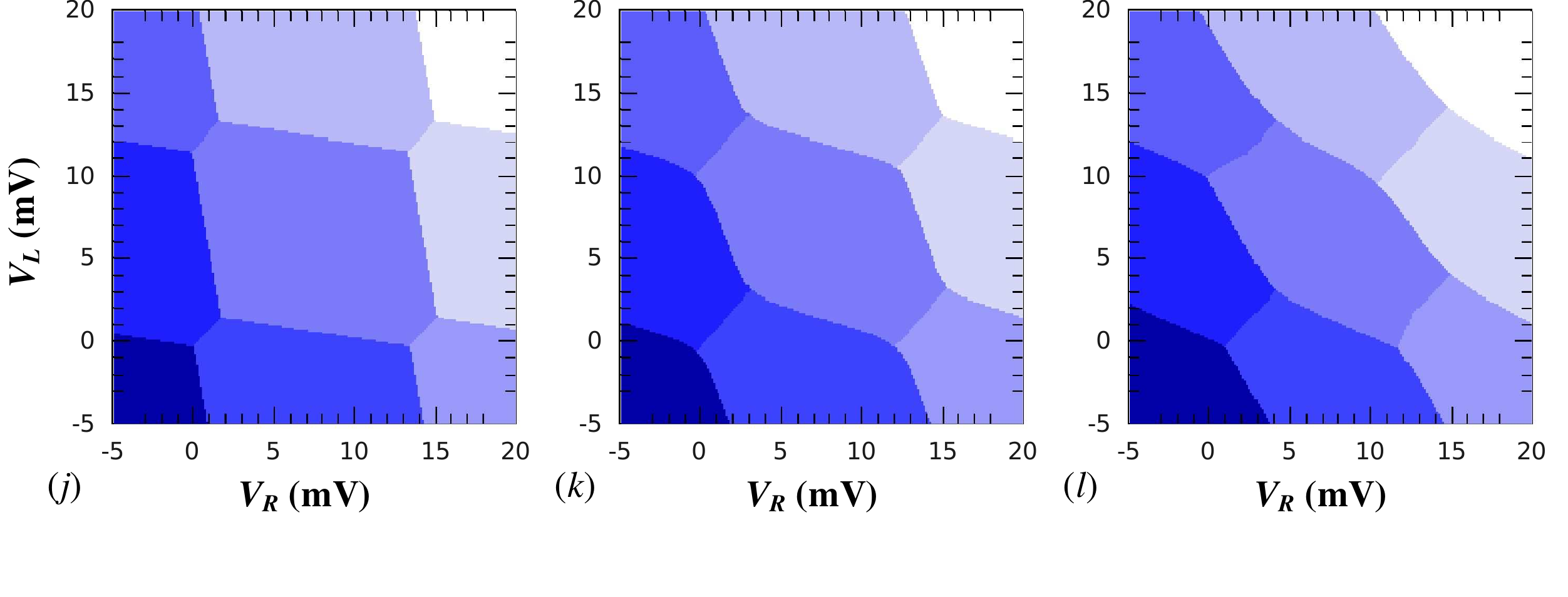}
\caption{(Color online) Comparison of calculated charge stability diagrams and the experiments on Si-MOS system.\cite{Lai.10} (a)-(c): experimental results reproduced from Fig.~2(a)-(c) of Ref.~\onlinecite{Lai.10}. The blue squares on the figures indicate the data range that we intend to fit. (d)-(f): Results from the Hubbard model, which fit panels (a), (b), and (c), respectively. Parameters: (d): $U_1=9.8$ meV, $U_2=11$ meV, $U_{12}=1.8$ meV, $t=0$, $\alpha_1=0.862$, $\alpha_2=0.842$. (e): $U_1=8.36$ meV, $U_2=9.38$ meV, $U_{12}=2.6$ meV, $t=0.3$ meV, $\alpha_1=0.791$, $\alpha_2=0.754$. (f): $U_1=6.53$ meV,  $U_2=7.33$ meV, $U_{12}=3.33$ meV, $t=0.4$ meV, $\alpha_1=0.710$, $\alpha_2=0.638$. All other parameters $J_e$, $J_p$, $J_{t1}$, $J_{t2}$ are zero. (g)-(i): Microscopic calculations [fitting (a)-(c)] with the biquadratic potential [Eq.~\eqref{biquadratic}]. Parameters: (g): $\hbar \omega_{1}=0.718$ meV, $\hbar \omega_{2}=0.905$ meV, $a=52.544$ nm, $\alpha_{1}=0.862$, $\alpha_{2}=0.842$. (h): $\hbar \omega_{1}=0.532$ meV, $\hbar \omega_{2}=0.671$ meV, $a=37.827$ nm, $\alpha_{1}=0.772$, $\alpha_{2}=0.732$. (i): $\hbar \omega_{1}=0.336$ meV, $\hbar \omega_{2}=0.421$ meV, $a=28.515$ nm, $\alpha_{1}=0.671$, $\alpha_{2}=0.608$. (j)-(l): Microscopic calculations with the Gaussian potential [Eq.~\eqref{gaussian}]. Parameters: (j): $V_{b}=37$ meV, $a_1=70$ nm, $a_2=56$ nm, $l_{d1}=62$ nm, $l_{d2}=58$ nm, $l_{bx}=15$ nm, $l_{by}=40$ nm, $\alpha_{1}=0.885$, $\alpha_{2}=0.866$. (k): $V_{b}=14$ meV, $a_1=36$ nm, $a_2=33$ nm, $l_{d1}=57$ nm, $l_{d2}=55$ nm, $l_{bx}=18$ nm, $l_{by}=25$ nm, $\alpha_{1}=0.983$, $\alpha_{2}=0.948$. (l): $V_{b}=2.2$ meV, $a_1=26$ nm, $a_2=20$ nm, $l_{d1}=55$ nm, $l_{d2}=51$ nm, $l_{bx}=30$ nm, $l_{by}=55$ nm, $\alpha_{1}=0.997$, $\alpha_{2}=0.965$.}
\label{fitLai}
\end{figure*}

\begin{figure*}
\centering
\includegraphics[width=13.5cm]{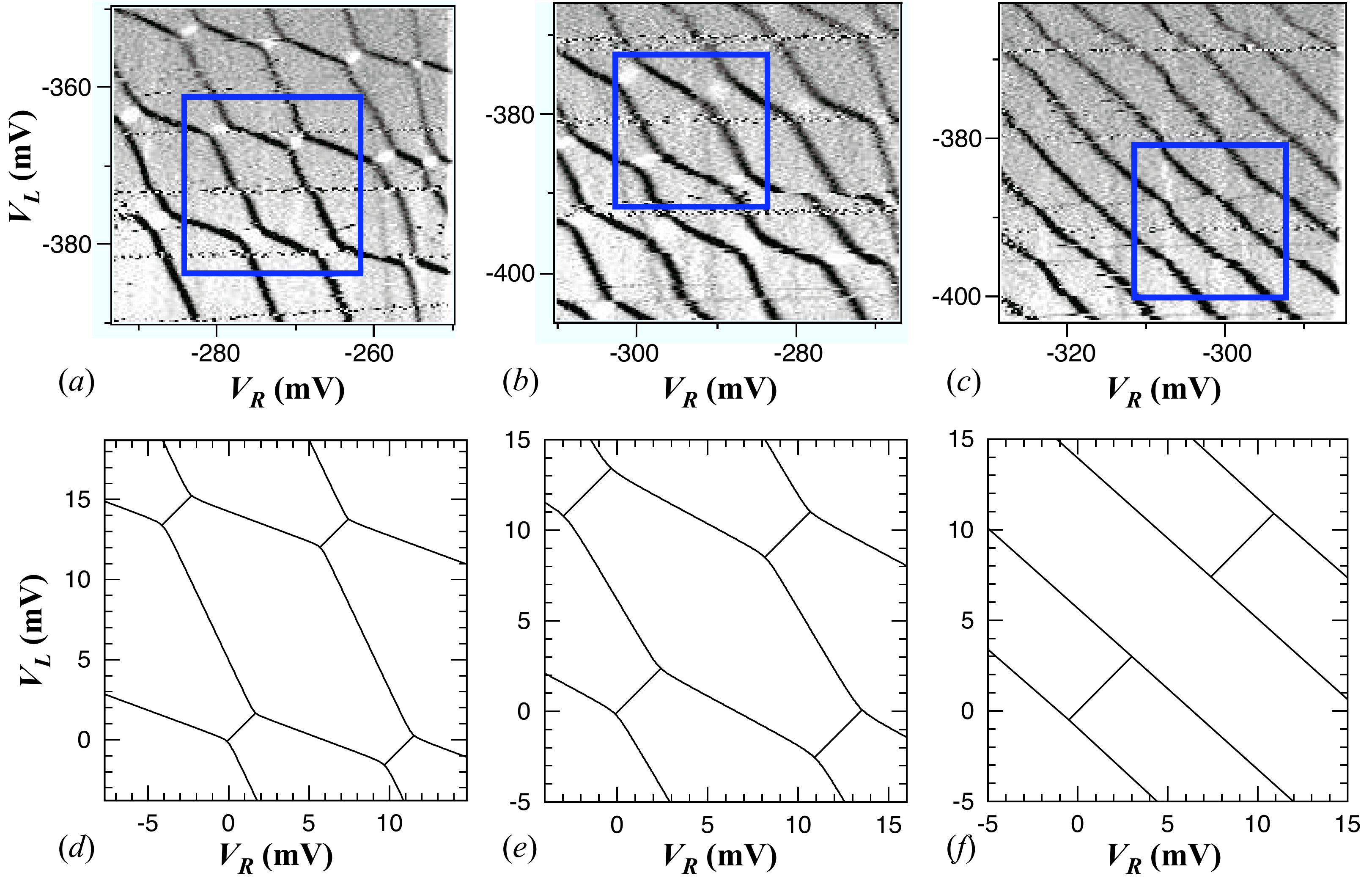}
\includegraphics[width=13.56cm]{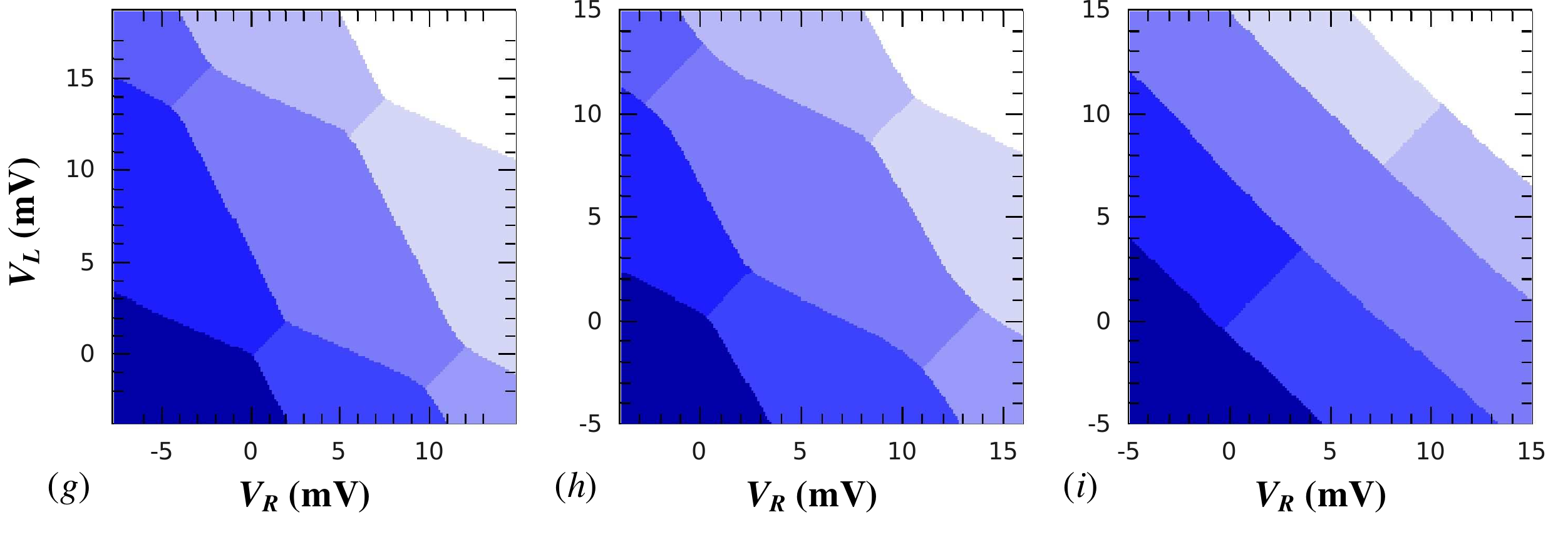}
\includegraphics[width=13.56cm]{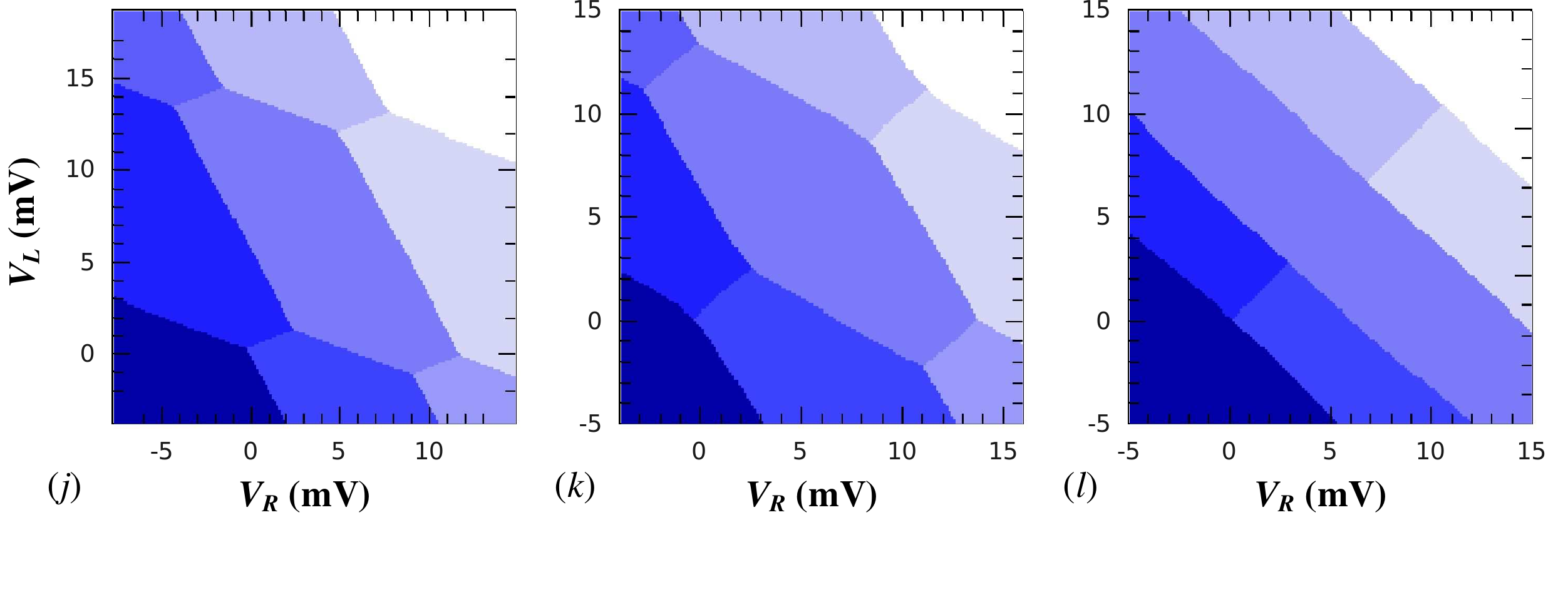}
\caption{(Color online) Comparison of calculated charge stability diagrams and experiments on Si/SiGe system.\cite{Simmons.09} (a)-(c): experimental results reproduced from Fig.~2(c)-(a) of Ref.~\onlinecite{Simmons.09}. (Note the reverse order) (d)-(f): Results from the Hubbard model, which fit the content enclosed by the blue frames in panels (a), (b), and (c), respectively. Parameters: (d): $U_1=8.8$ meV, $U_2=6.2$ meV, $U_{12}=1.6$ meV, $t=0.09$ meV, $\alpha_1=0.73$, $\alpha_2=0.68$. (e): $U_1=6.2$ meV, $U_2=6.1$ meV, $U_{12}=2.3$ meV, $t=0.12$ meV, $\alpha_1=0.653$, $\alpha_2=0.63$. (f): $U_1=4$ meV,  $U_2=4$ meV, $U_{12}=3.2$ meV, $t=0.5$ meV, $\alpha_1=0.515$, $\alpha_2=0.455$. All other parameters $J_e$, $J_p$, $J_{t1}$, $J_{t2}$ are zero. (g)-(i): Microscopic calculations with the biquadratic potential. Parameters: (g): $\hbar \omega_{1}=1.397$ meV, $\hbar \omega_{2}=0.723$ meV, $a=32.825$ nm, $\alpha_{1}=0.70$, $\alpha_{2}=0.66$. (h): $\hbar \omega_{1}=0.799$ meV, $\hbar \omega_{2}=0.773$ meV, $a=29.339$ nm, $\alpha_{1}=0.675$, $\alpha_{2}=0.635$. (i): $\hbar \omega_{1}=0.421$ meV, $\hbar \omega_{2}=0.421$ meV, $a=2.549$ nm, $\alpha_{1}=0.515$, $\alpha_{2}=0.48$. (j)-(l): Microscopic calculations with the Gaussian potential. Parameters: (j): $V_{b}=10$ meV, $a_1=25$ nm, $a_2=37$ nm, $l_{d1}=30$ nm, $l_{d2}=70$ nm, $l_{bx}=10$ nm, $l_{by}=20$ nm, $\alpha_{1}=0.95$, $\alpha_{2}=0.68$. (k): $V_{b}=9$ meV, $a_1=19$ nm, $a_2=18$ nm, $l_{d1}=40$ nm, $l_{d2}=43$ nm, $l_{bx}=25$ nm, $l_{by}=20$ nm, $\alpha_{1}=0.91$, $\alpha_{2}=0.86$. (l): $V_{b}=2$ meV, $a_1=4$ nm, $a_2=4$ nm, $l_{d1}=46$ nm, $l_{d2}=46$ nm, $l_{bx}=5$ nm, $l_{by}=10$ nm, $\alpha_{1}=0.98$, $\alpha_{2}=0.92$.}
\label{fitSimmons}
\end{figure*}

Here we discuss the experiments on Si-MOS devices.\cite{Lai.10} We start with the results from the Hubbard model, shown in Fig.~\ref{fitLai}(d)-(f). As discussed above, we assume that the capacitance model is valid in Fig.~\ref{fitLai}(a), meaning that in the generalized Hubbard model we need only retain parameters $U_1$, $U_2$, and $U_{12}$. In fact, in Ref.~\onlinecite{Lai.10} the charging energies, namely the on-site Coulomb interactions $U_1$ and $U_2$, are quantitatively given as $U_1=9.8$ meV and $U_2=11$ meV. We note that dot 1 is slightly larger than dot 2, different from the case with symmetric double dots discussed in Ref.~\onlinecite{Yang.11}. The gate voltages are transformed from the chemical potentials according to Eq.~\eqref{alphabeta}. With fixed $U_1=9.8$ meV and $U_2=11$ meV, the charge stability diagram in this case solely depends on $U_{12}$ and a fit to experiment can be readily obtained. We have found that $U_{12}=1.8$ meV as the appropriate parameter value and the result is shown in Fig.~\ref{fitLai}(d).

To fit Fig.~\ref{fitLai}(b) and (c) we switch on the most significant quantum parameter in our generalized Hubbard model: the tunnel coupling $t$. We note that the values of $U_1$, $U_2$ and $U_{12}$ need to be accordingly adjusted and we fix the ratio $U_1/U_2$ to be equal to the case of Fig.~\ref{fitLai}(d). The results are shown in Fig.~\ref{fitLai}(e) and (f). In Fig.~\ref{fitLai}(e), $U_{12}=2.6$ meV is intermediate compared to $U_1=8.36$ meV and $U_2=9.38$ meV which leads to the honeycomb shape, while a tunnel coupling of $t=0.3$ meV slightly rounds the phase boundaries near the triple points, in agreement with the experiment. In Fig.~\ref{fitLai}(f), $U_{12}=3.33$ meV is relatively large compared to $U_1=6.53$ meV and $U_2=7.33$ meV, while $t=0.4$ meV. This makes the phase boundaries almost like diagonal parallel lines and the rounding effect is substantial. We note that in the same $N$ block the boundaries separating different $(n_1,n_2)$ states are calculated according to their respective probabilities \cite{Yang.11} but not the differential conductance. Therefore the fading effect of those lines in the experiment due to quantum fluctuations is not seen in Fig.~\ref{fitLai}(f). The conductance calculations involve completely different physics where the excited states play a role which is well beyond the scope of our work where we restrict ourselves to the ground state physics and the associated charge stability diagram.

The results of the microscopic calculation with biquadratic potential [Eq.~\eqref{biquadratic}]  are shown in Fig.~\ref{fitLai}(g)-(i).  For Fig.~\ref{fitLai}(g) one sees excellent agreement with experiments and Hubbard model calculations. As far as Fig.~\ref{fitLai}(b) and (c) are concerned, the inter-dot coupling becomes important and Eqs.~\eqref{U1U2} and \eqref{U12} cannot be directly applied. However, as mentioned in the above section they do provide a reasonable initial guess which refines the searching in the parameter space. Fig.~\ref{fitLai}(h) and (i) present the results, and good agreement with both the experiments and the Hubbard model calculations are found. We note that the coefficients $(\alpha_1,\alpha_2)$ which transform from the $(\mu_2,\mu_1)$ to the $(V_R,V_L)$ planes are not chosen from Eq.~\eqref{alphabeta} since the capacitance model is invalid. Their values are provided in the caption of Fig.~\ref{fitLai}.

We also show the results from the Gaussian potential [Eq.~\eqref{gaussian}] in Fig.~\ref{fitLai}(j)-(l). Very good agreement with their respective counterparts is evident. The Gaussian potential contains seven parameters which can be independently tuned, as such it must be noted that the data fitting procedure involves a large degree of freedom of choosing parameters which is to some extent artificial, especially in our case where both the precise form of the confinement potential and the detailed electronic configuration in the system are unknown. We have made efforts to precisely reflect all reliable information that we currently have, however these restrictions still leave several free degrees of freedom. In the present work we have shown that our theory is capable of giving a quantitative explanation of experiments, with all reliably known restrictions incorporated. However an ultimate complete description of the experiment would require both improvement of the experimental technique and the first-principles Poisson-Schr\"odinger approach which treats the exact solution to the Schr\"odinger equation and a realistic confinement potential therein. Both of these are currently impractical to achieve since all the details of the actual potential confinement of the quantum dots are simply unknown for the experimental structures.

\subsection{Si/SiGe}

In this subsection we discuss the charge stability diagram and the tunnel coupling of the Si/SiGe devices. In Fig.~\ref{fitSimmons}(a)-(c) we replicate the experimental plot from Fig.~2(c)-(a) of Ref.~\onlinecite{Simmons.09}. The order is reversed with respect to Ref.~\onlinecite{Simmons.09} such that the tunnel coupling goes from weak to strong from panel (a) to (c). There are some noisy lines parallel to the $x$-axis ($V_R$) which slightly dislocate the diamonds. They are believed to result from impurities in the quantum dot\cite{Nguyen.11} and are not further discussed here. The rounded boundary lines near the triple points mean that even for the case of Fig.~\ref{fitSimmons}(a) one needs to take the tunnel coupling into account. Moreover, the asymmetry between the two dots is most pronounced in Fig.~\ref{fitSimmons}(a) but is substantially suppressed in Fig.~\ref{fitSimmons}(b) and (c), as can be seen directly by reflecting the figure with respect to the diagonal line $V_L-V_R={\rm const.}$. We suspect that it is due to the fundamental difference  in the way the Si-MOS and Si/SiGe devices are fabricated, therefore unlike the Si-MOS case where we fix the ratio $U_1/U_2$, in the Si/SiGe case we use independent $U_1$ and $U_2$ values that are appropriate for the given charge stability diagrams.

Fig.~\ref{fitSimmons}(d)-(f) present the results from the Hubbard model. In Fig.~\ref{fitSimmons}(d), the asymmetry is evident as $U_1=8.8$ meV and $U_2=6.2$ meV. A finite $t=0.09$ meV makes the boundary lines slightly rounded. In Fig.~\ref{fitSimmons}(e) the values of $U_1$ and $U_2$ are already quite similar ($U_1=6.2$ meV, $U_2=6.1$ meV), implying a rather symmetric honeycomb pattern. $t=0.12$ meV, which is larger than that of Fig.~\ref{fitSimmons}(d), makes the boundary line even more rounded. As stated in Ref.~\onlinecite{Simmons.09}, in Fig.~\ref{fitSimmons}(f) the double dot system effectively becomes a single, large quantum dot and our fitted parameters $U_1=U_2=4$ meV, and rather large values of $U_{12}=3.2$ meV and $t=0.5$ meV are consistent with this argument. We note that in Fig.~\ref{fitSimmons}(f) the $(0,2)$ and $(2,0)$ components of the stability diagram appear out of the range shown in the figure. This can be understood since, as the two dots effectively merge due to very low central barrier, the (1,1) state dominates most of the diagram and one must use substantially different chemical potentials on the two dots in order to achieve double occupancy on one but not the other of the two dots.

\begin{figure*}[t]
\centering
\includegraphics[width=11.5cm]{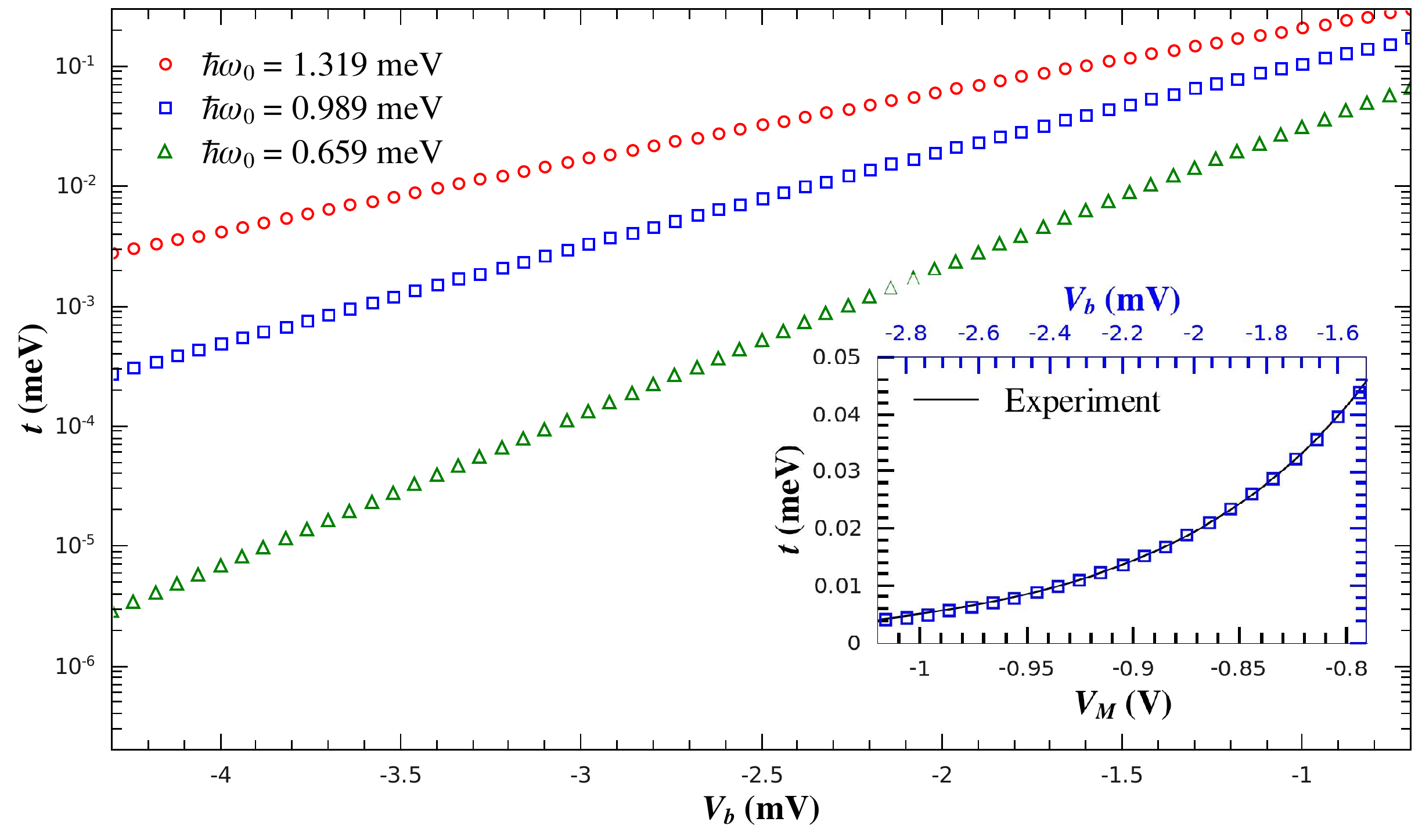}
\caption{(Color online) Tunnel coupling $t$ calculated from the biquadratic model potential as functions of the effective central gate voltage $V_b=h_{b}/(-e)$, plotted on the semi-log scale. Here $h_b$ is the height of the central potential barrier.  Since $t$ varies over the whole range of the charge stability diagram, we choose the chemical potentials at the center of (1,1) phase.  As the absolute value of $V_b$ decreases, $t$ increases approximately exponentially.  The harmonic oscillator energy level spacings of the single-dot confinements are fixed as $\hbar \omega_0=1.319$meV (red circles), $\hbar \omega_0=0.989$meV (blue squares), and $\hbar \omega_0=0.659$meV (green triangles). The inter-dot distance $2a=2\sqrt{2 h_{b} /m } /\omega_{0}$. Inset: Overlaying comparisons between $t$ extracted from experiment as a function of middle gate voltage $V_M$ (black solid line, bottom $x$-scale reproduced from Fig.~4 inset of Ref.~\onlinecite{Simmons.09}) and $t$ calculated from the microscopic model as a function of effective central gate voltage $V_b$ ( $\hbar \omega_0=0.989$meV, blue squares, top $x$-scale). The $y$-axis is shared by the two but note that the $x$-scales differ by a proportionality constant and a shift.}
\label{hoppingttt1}
\end{figure*}

We show the results of the microscopic calculation from the biquadratic model potential and the Gaussian potential in Fig.~\ref{fitSimmons}(g)-(i) and Fig.~\ref{fitSimmons}(j)-(l), respectively. The overall features are observed to be very similar as the parameters are tuned, but there are differences in the fine details. For example in Fig.~\ref{fitSimmons}(j) the boundaries between different $(n_1,n_2)$ states within the same $N$ block [e.g. $(1,1)$ and $(0,2)$] are tilted, and in Fig.~\ref{fitSimmons}(i) and (l) the spacing of the diagonal ``parallel lines'' are observed to be quite different from the experiments. This indicates that the model confinement potential is sometimes insufficient to mimic the true potential that electrons would feel in the quantum dot system. To resolve this problem one must calculate the electrostatic confinement potential according to the design of the devices using, for example, the Poisson equation. Since obtaining the precise connection between gates and the electron gas localized inside the quantum dots is difficult, one would not expect that this problem can be straightforwardly resolved along this line. Further study of this issue is needed.

The tunnel coupling $t$ is crucial in the double quantum dot system since it implies the exchange interaction, which is essential for many proposals of spin-based quantum computation.\cite{Loss.98,DiVincenzo.00,Levy.02,Laird.10} While the exchange interaction can be directly measured experimentally from the singlet/triplet level splitting,\cite{Lai.10} the tunnel coupling $t$ is of fundamental theoretical importance as it appears in the Hamiltonian matrix. As discussed in Ref.~\onlinecite{Yang.11}, there are at least two indirect ways to measure $t$: one is through the curvature of the boundary lines, as discussed above in the context of stability diagrams; the other is according to the width of the probability crossover near the
phase boundaries within a subspace of the charge stability diagram, for example the $(1,1)$ and $(0,2)$ probability crossing. The latter strategy has been used in Refs.~\onlinecite{Simmons.09} and \onlinecite{Hatano.05}: the crossover of $(n_1+1, n_2)$ and $(n_1, n_2+1)$ states leads to an obvious change in the tunneling current which is directly measured by the quantum point contact.  Then employing a two-level model (which is essentially a subspace of our more general model) one extracts the $t$ values. Fig.~4 of Ref.~\onlinecite{Simmons.09} shows  such a crossover of the occupancy and its inset shows the extracted values of $t$ as a function of the middle gate voltage $V_M$, whose fitting line is reproduced as the black solid line in the inset of Fig.~\ref{hoppingttt1}. As the absolute value of $V_M$ increases, the central potential barrier is raised and $t$ decreases approximately exponentially. 

Here we present a microscopic calculation which shows similar behavior and compare that to the experiments. We use the biquadratic model potential as it contains fewer free parameters and the inter-dot distance has a simple relation to the height of the potential barrier and the level spacing of the harmonic oscillator: $2a=2\sqrt{2 h_{b} /m } /\omega_{0}$. Since the precise microscopic parameters cannot be known from the experiments, we use a range of different harmonic oscillator level spacings, assuming the two dots to be symmetric, and study the tunnel coupling $t$ as a function of the height of the central potential barrier $h_b$. The results are shown in the main panel of Fig.~\ref{hoppingttt1}: the results of three values of harmonic oscillator level spacing are shown as different symbols with $\hbar \omega_0=1.319$meV (red circles), $\hbar \omega_0=0.989$meV (blue squares), and $\hbar \omega_0=0.659$meV (green triangles). Since the calculated value of  $t$ changes over the entire charge stability diagram, we focus on the center point of the (1,1) component of the charge stability diagram and compare all cases at that point. To facilitate the comparison to experiments we convert the central barrier height to the effective gate voltage $V_b=h_{b}/(-e)$. On the semi-log scale the data points show an approximately exponential behavior. The exponential behavior can again be understood from the local nature of the electron wave function: the overlap between two well localized wave function at a distance decreases rapidly as they become either further apart or more localized. 

We compare our results calculated at $\hbar \omega_0=0.989$ meV (blue squares) to experiments in the inset of Fig.~\ref{hoppingttt1}. Difficulties arise since, although the experimental middle gate voltage $V_M$ plays a similar role as the central potential barrier of the model potential $h_b$ in our calculation, the quantitative connection between the realistic confinement potential and our model potential is in general unknown. First, the realistic confinement potential depends on details of the fabrication and the geometry of the specific device, which is difficult to quantitatively describe. Second, the multi-electron multi-band feature in the real situation further complicates the problem. 
Therefore, in our situation, we shall only assume that $V_M$ and $V_b$ [$=h_b/(-e)$] are related linearly without quantifying the linear coefficients. In the inset of Fig.~\ref{hoppingttt1} the theoretical results are shown as points, utilizing the top $x$-scale, while the experimental fitting line is shown as the solid line, using the bottom $x$-scale. The $y$-scale is shared by both and the $x$-scales differ by a proportionality constant and a shift as discussed above. Caution must be taken when interpreting this figure: the perfect fit does not mean that the microscopic parameters used here are the correct ones for the corresponding experiment because of our many simplifying approximations. The purpose is only to demonstrate that our method is capable of producing the exponential behavior, assuming that $V_M$ and $V_b$ are related linearly. The precise connection between them is still uncertain. In fact each group of the three $\hbar\omega_0$ points can be appropriately rescaled to fit the experimental curve. Therefore we only argue that our theory produces the exponential behavior along with appropriate, albeit rough, simplifications, and leave the precise first-principles treatment for future studies. Such a study, which is well beyond the scope of our minimal model, will not only require intensive computational efforts but also necessitate much better characterization of the experimental dots than is currently available.

\section{Conclusion}\label{sec:conculsion}

In conclusion, we have applied the generalized two-site Hubbard model approach previously introduced in the GaAs symmetric double dot system to the Si asymmetric double dot system. We have calculated the charge stability diagrams from the generalized Hubbard model and the microscopic theory with two different models of confinement potentials (biquadratic and Gaussian forms), and have found qualitative and quantitative agreement between them and with the experiments on Si-MOS and Si/SiGe devices. We have explained in detail the simplifications that we have made to apply our theory, which involves the reduction to the effective two-electron regime\cite{long} and approximation to the gate voltages. We have shown that our proposed theoretical Hubbard model approach, along with appropriate approximations, provides state-of-the-art quantitative description to the silicon double quantum dot system. In addition, we have studied in particular the tunnel coupling using the microscopic calculation with biquadratic model potential and have shown that the experimentally observed exponentially decaying behavior with the central potential barrier height is consistent with our theory. 

The qualitative success of our minimal theoretical model should motivate further investigations. First, the full understanding of the multi-electron nature of silicon quantum dot devices requires a multi-band calculation.\cite{Hu.01} In our work we have qualitatively argued on physical grounds that the multi-electron effect imposes simply a renormalization of Hubbard parameters in the generalized one-band effective Hubbard model,\cite{long} but a quantitative assessment is lacking. It is important to understand in-depth how the Hubbard parameters change as one translates between different multi-electron regimes, in particular those corresponding to quantum fluctuations which may have non-trivial dependence on the configuration of the base Fermi sea. Second, a first-principles Poisson-Schr\"odinger theory would be helpful in determining the microscopic confinement potential. The deficiency of current experiments is that the confinement potential is subject to fabrication-induced disorder and it is hard to extract the precise form of confinement potential reliably. Beyond the Hund-Mulliken approximation, one can in principle obtain exact solutions to the Schr\"odinger equation for any arbitrary confinement potential, although for some calculations the cost is expensive. However a complete quantitative understanding will be impossible unless one acquires knowledge beyond the model potential and the effective two-electron regime.
Some of the above problems are intractable for current technologies. However, these warrant further study. A key point to appreciate in this context is the fact that no two sets of double dots are identical from the perspective of microscopic details even when they are fabricated from the same underlying two-dimensional electron structures using the same nominal lithographic protocols.   This is due to many reasons, but a primary one is the invariable presence of unintentional (and therefore, unknown) background disorder in the host semiconductor, particularly at the interface where the dots reside.  It may therefore be useless trying to go much beyond our effective minimal model at this stage of materials development in solid state spin qubits since no generic quantitative  theory  is feasible until the characterization of the quantum dots and their underlying materials science become much more advanced and reproducible than they currently are.  We believe that our minimal model is perhaps the most practical quantum generalization of the classical capacitance model that is feasible at this stage of development of solid state spin qubits in semiconductors.

\section*{Acknowledgements}
We thank M. A. Eriksson, J. P. Kestner and Q. Li for helpful discussions.
This work is supported by LPS-CMTC.

\end{document}